\RequirePackage{ifpdf}
\documentclass{PoS}
\usepackage[authoryear]{natbib}
\bibpunct{(}{)}{;}{a}{}{,}

\newcommand{\araa}{Annual Reviews of Astronomy \& Astrophysics}
\newcommand{\apss}{Astrophysics and Space Science}
\newcommand{\apj}{The Astrophysical Journal}
\newcommand{\aj}{The Astronomical Journal}
\newcommand{\mnras}{Monthly Notices of the Royal Astronomical Society}
\newcommand{\aap}{Astronomy \& Astrophysics}

\title{Tomography of Galactic star-forming regions and spiral arms with the Square Kilometer Array}

\ShortTitle{Tomography with the SKA}

\author{
Laurent Loinard$^1$,
Mark Thompson$^2$, 
Melvin Hoare$^3$, 
Huib Jan van Langevelde$^4$,
Simon Ellingsen$^5$,
Andreas Brunthaler$^6$,
Jan Forbrich$^7$,
Kazi L.J.\ Rygl$^8$,
Luis F.\ Rodr\'{\i}guez$^1$,
Amy J.\ Mioduszewski$^9$,
Rosa M.\ Torres-L\'opez$^{10}$,
Sergio A.\ Dzib$^6$,
Gisela N.\ Ortiz-Le\'on$^1$,
Tyler L. Bourke$^{11}$,
James A. Green$^{11}$\\
$^1$CRyA-UNAM Morelia~~
$^2$University of Hertfordshire~~
$^3$Leeds University~~
$^4$JIVE/Sterrewacht Leiden~~
$^5$University of Tasmania~~
$^6$MPIfR Bonn~~
$^7$University of Vienna~~
$^8$ESA-ESTEC Noordwijk~~
$^9$NRAO Socorro~~
$^{10}$Universidad de Guadalajara~~
$^{11}$SKA Organisation
\\
E-mail: \email{l.loinard@crya.unam.mx}
}

\abstract{Very Long Baseline Interferometry (VLBI) at radio wavelengths can
provide astrometry accurate to 10 micro-arcseconds or better (i.e.\ better
than the target GAIA accuracy) without being limited by dust obscuration.
This means that unlike GAIA, VLBI can be applied to star-forming regions
independently of their internal and line-of-sight extinction. Low-mass
young stellar objects (particularly T Tauri stars) are often non-thermal
compact radio emitters, ideal for astrometric VLBI radio continuum
experiments. Existing observations for nearby regions (e.g.\ Taurus,
Ophiuchus, or Orion) demonstrate that VLBI astrometry of such active T
Tauri stars enables the reconstruction of both the regions' 3D structure
(through parallax measurements) and their internal kinematics (through
proper motions, combined with radial velocities).  The extraordinary
sensitivity of the SKA telescope will enable similar {\em tomographic
mappings} to be extended to regions located several kpc from Earth, in
particular to nearby spiral arm segments. This will have important
implications for Galactic science, galactic dynamics and spiral structure
theories.}

\FullConference{Advancing Astrophysics with the Square Kilometre Array,\\
		June 8-13, 2014\\
		Giardini Naxos, Sicily, Italy }

\begin{document}
\makeatletter
\setbox\@firstaubox\hbox{\small Laurent Loinard}
\makeatother

\section{Introduction}

Very Long Baseline Interferometry at radio frequencies
\citep[e.g.][]{thompson2007} can readily provide astrometric accuracies of
order 10 micro-arcseconds ($\mu$as) or better, for compact radio sources of
sufficient brightness ($T_b$ $\approx$ 10$^7$ K). This enables the
determination of trigonometric parallaxes with an accuracy of a few percent
for any source within a few kpc (for instance, a 10 $\mu$as astrometric
accuracy translates to a 2\% parallax accuracy at 2 kpc). Similarly, proper
motions can be measured to an accuracy of 0.2 km s$^{-1}$ in one year at a
distance of 2 kpc. As a consequence, {\bf radio VLBI measurements have the
potential to provide distances accurate to better than a few percent, and
tangential velocities accurate to better than a few tenths of a km s$^{-1}$
for any source within a few kpc} \citep{reidhonma2014}. The GAIA space
mission is expected to deliver similar results, and will do so for many
millions of stars \cite[]{deBruijne2012}. However, radio VLBI observations
have the distinct advantage that they are not affected by dust extinction.
Thus, they can be used to complement GAIA for sources that are either
deeply embedded within dusty regions, or located behind a large column of
line-of-sight dust (or both). Star-forming regions are clearly in this
situation, and are therefore prime targets for VLBI astrometric
observations.

The astrometric potential of radio VLBI observations can only be realized
if the intended targets are detectable. As mentioned earlier, this requires
the brightness temperature to be of order 10$^7$ K \cite[]{thompson2007}, and
effectively restricts the pool of potential targets to non-thermal
emitters.\footnote{For instance, a compact H{\sc ii} region with a
brightness temperature of order 10$^4$ K is undetectable with a VLBI
array.} In star-forming regions, there are two classes of sources that meet
the brightness criteria: masers and chromospherically active young stars.
Masers (hydroxyl, water, methanol, and to a lesser extent silicon monoxide)
are commonly found in high-mass star-forming regions \citep{huib2012} and
can be extremely bright. This makes them ideal tracers to map the
distribution of high-mass star-forming regions across the Milky Way
\citep{reid2014,green2015}. Chromospherically active young stars, on the
other hand, are often radio sources thanks to the gyration of relativistic
electrons in their strong surface magnetic fields \citep{dulk1985}. This
results in continuum cyclotron, gyrosynchrotron, or synchrotron emission
depending on the energy of the gyrating electrons. This emission is
normally confined to regions extending only a few stellar radii around the
stars, and therefore remains very compact even in the nearest star-forming
regions (4 $R_{\odot}$ $\equiv$ 50 $\mu$as at 300 pc). We note that for
this mechanism to operate, a strong magnetic field (kGauss) must exist, and
this normally requires the dynamo process to operate within the young star.
This, in turns, requires the star to be convective, so only {\bf low-mass}
young stars are expected non-thermal radio emitters. There are, however,
some exceptions to this rule \citep{loinard2008,dzib2010}.

\begin{figure}
\includegraphics[width=0.99\textwidth]{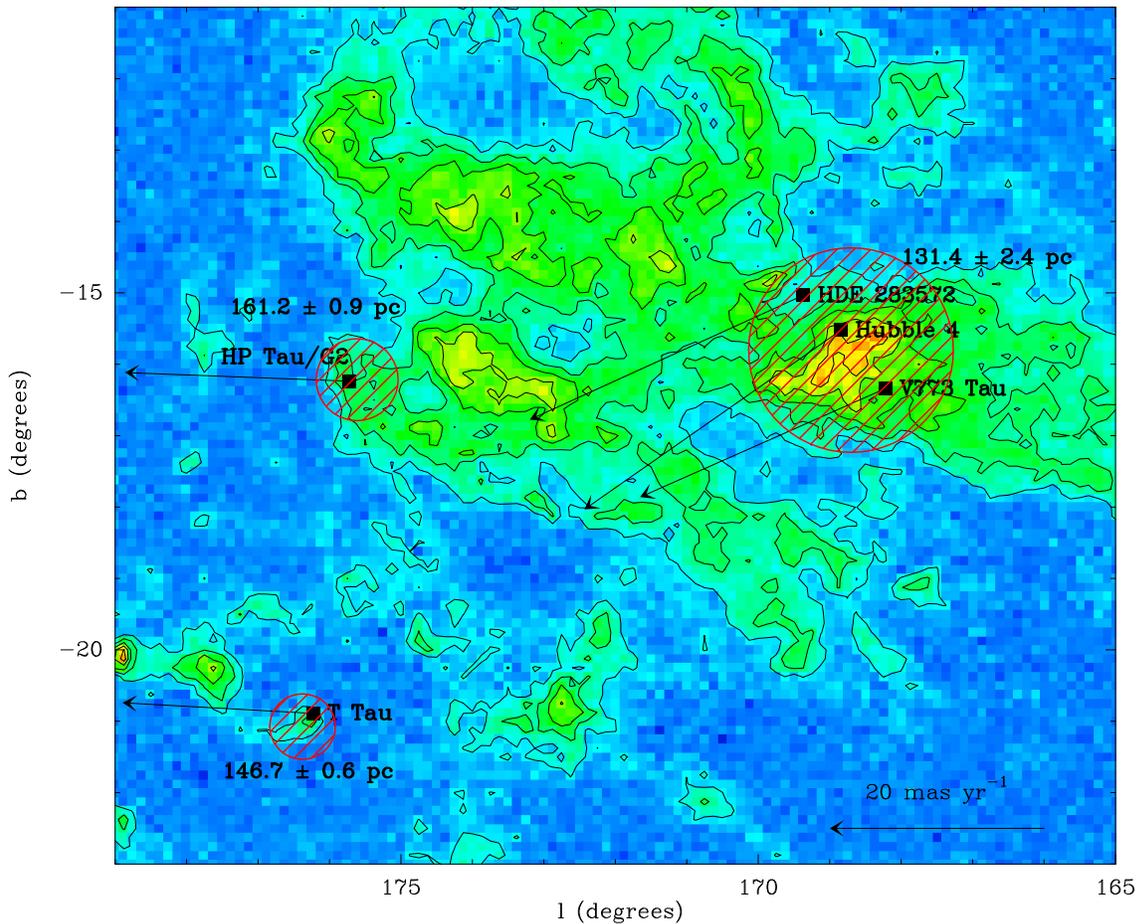}
\caption{Positions, distances, and proper motions of Hubble 4, HDE 283572, V773 Tau, T Tau, and HP Tau/G2 superposed on the CO(1-0) map of Taurus from \cite{dame2001}. Adapted from \cite{loinard2007,torres2007,torres2009,torres2012}.}
\label{fig:taurus}
\end{figure}

Since masers and magnetically active low-mass young stars can both be used
for VLBI astrometric observations, it is useful to consider their relative
merits. Masers are typically much brighter and can therefore be detected
much farther than magnetically active young stars. On the other hand,
active young stars are much more ubiquitous. First, they do exist in
regions of low, intermediate, and high-mass star formation, whereas bright,
steady masers are largely restricted to high-mass regions. A limited number
of masers have been identified in association with low-mass stars but they
are typically of lower luminosity \citep{kalenskii2010} than those found
toward high-mass star formation regions. Second, tens or hundreds of
magnetically active low-mass young stars might be present in a high-mass
star-forming region where only a few strong maser spots exist. For
instance, only a few maser sources are known in the Orion region
\citep{gaume1998} where we have recently detected well over 50 magnetically
active young stars in moderately deep VLBA observations (M.\ Kounkel et
al., in preparation). A very interesting application of this ubiquity is
that distances and proper motions can be measured to stars distributed
across a given region, and the three-dimensional structure and kinematics
of the region can be reconstructed. This has recently be exemplified in the
case of the Taurus star-forming region \citep{torres2007,torres2009,torres2012} thanks to multi-source VLBA observations. As Figure 1 shows, the young stars located to the west of the Taurus regions are at a common distance of 130 pc, and share similar proper motions. Measurements from the literature show that they also have similar radial velocities. The stars to the east and south of the complex, on the other hand, are significantly farther (145-160 pc) and have different proper motions and radial velocities. Although based on a very limited number of targets, this example demonstrates that multi-source VLBI observations enable {\it tomographic mappings} of individual star-forming regions. In this contribution, we build the case for extending such observations using VLBI with the SKA to regions farther from Earth, and particularly to nearby segments of spiral arms. 

\section{Scientific Rationale and synergy with other projects}

Thanks to its enormous collecting area, the SKA will provide unprecedented
sensitivity for radio continuum observations. As will be shown below, this
will enable tomographic mappings with unprecedented accuracy of nearby
regions such as Orion in a fairly modest amount of telescope time, even
during the SKA1-MID phase. Such detailed reconstruction of the 3D structure
and kinematics of individual star-forming regions will provide important
tests for models of star-formation and early stellar evolution. For
instance, a complete 6D (3 spatial + 3 velocity components) description of
the young stellar population in a given cluster could be used to test
theoretical models such as those proposed by Hartmann \& Burkert (2007) for
Orion. Similarly, very accurate HR diagrams made possible by accurate
distances could be used to constrain pre-main sequence evolutionary models
\citep[see][for details]{loinard2011}. It will also be interesting the
characterize the magnetospheric activity on young stars \citep{jan2011} by
statistically examining the relation between evolutionary stage, mass, and
magnetic activity. 

For a full description of the 3D structure of star-forming regions, one has
to consider, apart from the young stars and star-forming regions, also the
molecular cloud. It impossible to measure parallaxes to extended structures
such as clouds, but one can unravel the cloud through its dust content,
assuming a ratio between the gas and the dust. The advantage of the dust is
that we can measure its optically thin emission in the infrared to
sub-millimeter and at the same time that dust will redden the stellar light
in the optical and the near-infrared. Stars with accurate distances and
stellar types can thus be used as pinpoints to reconstruct the cloud
structure by measuring their reddening (for example Knude 2009). As clouds
can have extents of several tens of parsecs, the distance accuracies have
to be high, better than $\sim$10 pc, to obtain a detailed structure of the
cloud.  

Imposing a 2\% parallax accuracy requirement, GAIA will deliver astrometry of B-G dwarfs out to 3--1\,kpc (for B dwarfs) and 850--200 pc (G dwarfs) for visual extinctions of 1 to 10 mag, respectively. Later, but more common spectral types, such as M stars, have their astrometry limited to the very nearby stars only (200--50 pc for $A_V$ of 1-10 magnitudes). Herschel maps show that the nearby low-mass clouds have most of their extent at $A_V$ < 10 mag (see e.g., Peretto et al. 2012, Rygl et al.\ 2013, Kirk et al.\ 2013). Hence, combining VLBI and GAIA astrometries, we can measure both the low-density and the high-density extent of nearby molecular clouds, and have a complete 3D picture of the stars {\em and} the cloud, and the kinematics of the young and main sequence stars of stellar complexes. With early type stars (O-A) such studies could be extended out to the nearby segments of spiral arms.

With the full SKA2, VLBI tomographic mappings could be extended to the nearest spiral arm segments at 2--3 kpc (see Section 4.2). Spiral arms are likely to have a width of at least several hundred parsecs (several times the size of a massive star-forming region like Orion), so a parallax accuracy of order 1\% at 2.5 kpc would easily reveal the depth of a spiral arm segment. Thus, the tomography proposed here would reveal the inner structure of Galactic spiral arms in unprecedented details, as well as the 3D motions of the young stars along and across the arm. This could be directly compared with the predictions of theoretical models \cite[e.g.][]{bertin1996}, and would represent a major contribution to the study of the dynamics of galactic disks in general. It is important to stress that the level of details that would be achieved through this type of observations could not be obtained in any external galaxy for the foreseeable future, nor in our own Galaxy with GAIA due to a combination of crowding and dust obscuration. 

There is a clear synergy between the tomographic mappings proposed here and
the maser and pulsar astrometric observations described in
\cite{green2015} and \cite{han2015}.

\section{VLBI observations incorporating the SKA and strategies for astrometry}

In all that follows, we will assume that the observations are made at a frequency of order 6 GHz, i.e.\ at the lower frequency end of SKA Band 5. We consider a VLBI array consisting of 70\% of the SKA1-MID collecting area phased-up, and a further 10 other stations with characteristics similar to those of an individual SKA1-MID antenna. Assuming a 1~GHz bandwidth  (the sum of two orthogonal polarizations; implying a data rate of 4 Gbps per antenna), we obtain an rms per baseline (phased SKA1-MID to single antenna) in the continuum of 220~$\mu$Jy in 60 seconds and 97~$\mu$Jy in 5 minutes.  We estimate the image sensitivity by dividing the baseline sensitivity by $\sqrt{N_{ant} -1}$ and applying no weighting, since baselines to the phased SKA1-MID are so much more sensitive than those between the rest of the antennas in the array. The corresponding continuum image sensitivities (1$\sigma$) for 1 and 5 minutes are 97 and 31~$\mu$Jy respectively. 

A crucial ingredient for accurate astrometric measurements with the SKA is
the presence of nearby calibrators, ideally within the primary beam of the
SKA antennas, so that with multiple beams from the phased SKA1-MID we can
simultaneously observe both the targets and the background quasar.  The VLA
was used by \cite{fomalont1991} to characterize the faint source population
at 5 GHz and found that $N$, the number of sources per square arc minute
stronger than flux density $S$ (where $S$ is in $\mu$Jy) is given by \[
N(S) = (23.2\pm2.8) S^{-1.18\pm0.19}. \]The fraction of faint 5~GHz sources
which are compact on VLBI scales has (to our knowledge) not been measured
but it is safe to assume that it is higher than 20\% (the fraction of faint
source population detected in deep VLA observations at 1.4~GHz that are
compact on VLBI angular scales \citep{garrett2005}). Taking all this into
account, we predict 3 sources in the primary beam stronger than 150~$\mu$Jy
and perhaps one source stronger than 500~$\mu$Jy.  

The astrometry precision for a single epoch VLBI observation is given by \[ \theta_{ast} = \frac{\theta_{beam}}{2 \times \mbox{SNR} \sqrt{N_{cal}}},  \] where $N_{cal}$ is the number of calibrators. For a maximum baseline length of 8000~km we have a synthesized beam of 1.2~mas (3.1~mas for a 3000~km baseline).  To achieve an astrometric accuracy of 5~$\mu$as after 4 epochs (which would yield distances accurate to 1\% at 2~kpc), we will need a per epoch accuracy of around 10~$\mu$as (including systematic components of astrometric calibration).  We see from above that we need the product $2 \times \mbox{SNR} \times \sqrt{Ncal}$ to be approximately 120 if we have an 8000~km baseline length or 310 if the maximum baseline length is shorter. For 150~$\mu$Jy calibrators we then need an SNR of around 30 (3 in-beam calibrators), which implies on-source time per epoch of 4 hours per source.  For a 500~$\mu$Jy calibrator we need an SNR of around 60 (only 1 in-beam calibrator), which implies an on-source time per epoch of 1 hour.

\section{Staged approach to tomographic mappings of star-forming regions with the SKA}

\subsection{SKA1-MID}

It is clear from the previous section that the SKA will have the potential of carrying out detailed tomographic observations of regions of star-formation located at distances of up to several kpc. We propose to realize this potential through a staged approach. In an initial phase, we would focus on a nearby region (e.g.\ the Orion Nebula Cluster -- ONC) to demonstrate the capabilities of VLBI astrometric observations with the SKA1-MID. We would observe at least 200 young stars distributed across the complex (we know from VLA and VLBI observations that this number of non-thermal targets is present within the region). This would require observing about 20 primary fields of view distributed across the region, and ideally the capability of targeting simultaneously 10 to 20 fields within each primary field. With the current SKA1-MID baseline design of four VLBI beams per primary beam, and restricting ourselves to one calibrator and three targets per primary beam, we would need to observe each primary field at least 3 times. Assuming further that we focus on fields where calibrators brighter than 500~$\mu$Jy exist, we would need 4 (epochs) $\times$ 20 (primary fields) $\times$ 3 passes per field $\times$ 1 hour (per field) = 240 hours. Including a 20\% overhead brings the total required time for this project to 300 hours. If the capability of 10 to 20 VLBI beams was implemented in phase 1 this project would take only 100 hrs.

\subsection{SKA2}

If the observations with SKA1-MID are successful, we would move on to a more ambitious project utilizing the full SKA2. Taking advantage of the order of magnitude improvement in sensitivity, we would be able to carry out similar tomographic observations up to 2-3 kpc, i.e. the distance of the nearest spiral arm segments. For time estimate purposes, we will consider the mapping of a representative portion of a spiral arm segments of length 250 pc, and height 60 pc located at about 2.5 kpc of the Earth. This corresponds to a solid angle of 32,900 square arcminutes, or 180 SKA fields of view. We would, of course, only image the portions of the fields that contain young stellar objects or calibrators, thereby alleviating the computational requirements of the experiments. Assuming again 4 epochs, one hour per epoch, and 20\% overheads, we obtain a total of 1,000 hours to obtain a complete tomographic mapping of a nearby spiral arm segment.

\section{Conclusions and specific requirements}

Thanks to its extraordinary collecting area, the SKA will be the premier instrument for radio frequency astrometry. In this contribution, we make the case for using that capability to obtain tomographic mapping of star-forming regions and spiral arm segments within several kpc of the Earth. This would represent a unique contribution to galactic dynamics, and would have important implications for spiral structure theories. 

The general requirements to reach the necessary astrometric accuracy are
given in detail in \cite{green2015} and \cite{paragi2015}. Here we emphasize that to carry out the tomographic observations presented here, at least a few tens of targets would have to be observed in a single primary beam of the SKA1-MID antennas. This would include both young stellar sources (targets) and background quasars (calibrators). Without this capability, the time needed to carry out the observations would become prohibitive.

\setlength{\bibsep}{0.0pt}
\bibliographystyle{apj}

\begin{thebibliography}{99}

\bibitem[Bartkiewicz \& van Langevelde (2012)]{huib2012} Bartkiewicz, A., \&
van Langevelde, H.~J.\ 2012, IAU Symposium, 287, 117 

\bibitem[Bertin \& Lin (1996)]{bertin1996} Bertin, G., \& Lin, C.~C.\ 1996,
Spiral structure in galaxies a density wave theory, Publisher: Cambridge,
MA MIT Press, 1996 

\bibitem[de Bruijne (2012)]{deBruijne2012} de Bruijne, J.~H.~J.\ 2012,
\apss, 341, 31 

\bibitem[Dame et al.\ (2001)]{dame2001} Dame, T.~M., Hartmann, D., \&
Thaddeus, P.\ 2001, \apj, 547, 792

\bibitem[Dulk (1985)]{dulk1985} Dulk, G.~A.\ 1985, \araa, 23, 169 

\bibitem[Dzib et al.\ (2010)]{dzib2010} Dzib, S., Loinard, L., Mioduszewski,
A.~J., et al.\ 2010, \apj, 718, 610 

\bibitem[Fomalont et  al.\ (1991)]{fomalont1991} Fomalont, E.~B., Windhorst,
R.~A., Kristian, J.~A., \& Kellerman, K.~I.\ 1991, \aj, 102, 1258 

\bibitem[Forbrich et al.\ (2011)]{jan2011} Forbrich, J., Wolk, S.~J.,
G{\"u}del, M., et al.\ 2011, 16th Cambridge Workshop on Cool Stars, Stellar
Systems, and the Sun, 448, 455 

\bibitem[Garrett (2005)]{garrett2005} Garrett, M.~A.\ 2005, EAS
Publications Series, 15, 73 

\bibitem[Gaume et al.\ (1998)]{gaume1998} Gaume, R.~A., Wilson, T.~L.,
Vrba, F.~J., Johnston, K.~J., \& Schmid-Burgk, J.\ 1998, \apj, 493, 940 

\bibitem[Green et al.\ (2015)]{green2015} Green, J., et al.\ 2015, in
proceedings of "Advancing Astrophysics with the SKA" PoS(AASKA2014)119

\bibitem[Han et al.\ (2015)]{han2015} Han, J., et al.\ 2015, in proceedings
of "Advancing Astrophysics with the SKA" PoS(AASKA2014)041

\bibitem[Hartmann \& Burkert (2007)]{hartmann2007} Hartmann, L., \&
Burkert, A.\ 2007, \apj, 654, 988 

\bibitem[Kalenskii et al.\ (2010)]{kalenskii2010} Kalenskii, S.~V.,
Johansson, L.~E.~B., Bergman, P., et al.\ 2010, \mnras, 405, 613 

\bibitem[Kirk et al.\ (2013)]{2013MNRAS.432.1424K} Kirk, J.~M., Ward-Thompson, D., Palmeirim, P., et al.\ 2013, \mnras, 432, 1424 

\bibitem[Knude (2009)]{2009IAUS..254P..35K} Knude, J.\ 2009, IAU Symposium,
254, 35

\bibitem[Loinard et al.\ (2011)]{loinard2011} Loinard, L., Mioduszewski,
A.~J., Torres, R.~M., et al.\ 2011, Revista Mexicana de Astronomia y
Astrofisica Conference Series, 40, 205 

\bibitem[Loinard et al.\ (2007)]{loinard2007} Loinard, L., Torres, R.~M.,
Mioduszewski, A.~J., et al.\ 2007, \apj, 671, 546 

\bibitem[Loinard et al.\ (2008)]{loinard2008} Loinard, L., Torres, R.~M.,
Mioduszewski, A.~J., \& Rodr{\'{\i}}guez, L.~F.\ 2008, \apj, 675, L29 

\bibitem[Paragi (2015)]{paragi2015} Paragi, Z., et al.\ 2015, in
proceedings of "Advancing Astrophysics with the SKA" PoS(AASKA2014)143

\bibitem[Peretto et al.\ (2012)]{2012A&A...541A..63P} Peretto, N., Andr{\'e}, P., K{\"o}nyves, V., et al.\ 2012, \aap, 541, AA63 

\bibitem[Reid \& Honma(2014)]{reidhonma2014} Reid, M.~J., \& Honma, M.\
2014, \araa, 52, 339 

\bibitem[Reid et al.\ (2014)]{reid2014} Reid, M.~J., Menten, K.~M.,
Brunthaler, A., et al.\ 2014, \apj, 783, 130 

\bibitem[Rygl et al.\ (2013)]{2013A&A...549L...1R} Rygl, K.~L.~J.,
Benedettini, M., Schisano, E., et al.\ 2013, \aap, 549, LL1 

\bibitem[Thompson et al.\ (2007)]{thompson2007} Thompson, A.~R., Moran,
J.~M., \& Swenson, G.~W.\ 2007, Interferometry and Synthesis in Radio
Astronomy, by A.R.~Thomspon, J.M.~Moran, and G.W.~Swenson.~John Wiley \&
Sons, 2007.

\bibitem[Torres et al.\ (2007)]{torres2007} Torres, R.~M., Loinard, L.,
Mioduszewski, A.~J., \& Rodr{\'{\i}}guez, L.~F.\ 2007, \apj, 671, 1813 

\bibitem[Torres et al.\ (2009)]{torres2009} Torres, R.~M., Loinard, L.,
Mioduszewski, A.~J., \& Rodr{\'{\i}}guez, L.~F.\ 2009, \apj, 698, 242 

\bibitem[Torres et al.\ (2012)]{torres2012} Torres, R.~M., Loinard, L.,
Mioduszewski, A.~J., et al.\ 2012, \apj, 747, 18 

\end{thebibliography}

\end{document}